\documentclass[sigconf, screen]{acmart}
\pdfoutput=1 
\AtBeginDocument{%
  \providecommand\BibTeX{{%
    \normalfont B\kern-0.5em{\scshape i\kern-0.25em b}\kern-0.8em\TeX}}}

\copyrightyear{2023}
\acmYear{2023}
\setcopyright{rightsretained}
\acmConference[CHI EA '23]{Extended Abstracts of the 2023 CHI Conference on Human Factors in Computing Systems}{April 23--28, 2023}{Hamburg, Germany}
\acmBooktitle{Extended Abstracts of the 2023 CHI Conference on Human Factors in Computing Systems (CHI EA '23), April 23--28, 2023, Hamburg, Germany}\acmDOI{10.1145/3544549.3585694}
\acmISBN{978-1-4503-9422-2/23/04}

\begin{document}

\title[The Human Behind the Data]{The Human Behind the Data: Reflections from an Ongoing Co-Design and Deployment of a Data-Navigation Interface for Front-Line Emergency Housing Shelter Staff}


\author{Teale W. Masrani}
\email{teale.masrani2@ucalgary.ca}
\orcid {0000-0001-7749-1379}
\affiliation{%
  \institution{Department of Computer Science, University of Calgary}
   \city{Calgary}
   \country{Canada}
}

\author{Helen Ai He}
\email{helen.he1@ucalgary.ca}
\orcid{0000-0001-8681-2381}
\affiliation{%
  \institution{Department of Computer Science, University of Calgary}
   \city{Calgary}
   \country{Canada}
}

\author{Geoffrey Messier}
\email{gmessier@ucalgary.ca}
\orcid {0000-0002-9825-3238}
\affiliation{%
  \institution{Electrical and Software Engineering, University of Calgary}
   \city{Calgary}
   \country{Canada}
}

\renewcommand{\shortauthors}{Masrani, et al.}



\begin{abstract}
On any night in Canada, at least 35,000 individuals experience homelessness. These individuals use emergency shelters to transition out of homelessness and into permanent housing. We designed and deployed a technology to support front-line staff at the largest emergency housing shelter in Calgary, Canada. Over a period of five months in 2022, we worked closely with front-line staff to co-design an interface for supporting a holistic understanding of client context and facilitating decision-making. The tool is currently in-use and our collaboration is ongoing. In this paper, we reflect on preliminary findings regarding the second iteration of the tool. We find that supporting shelter staff in understanding \textit{the human behind the data} was a critical component of design. This work contributes to literature on how data tools may be integrated into homeless shelters in a way that aligns with shelters' values.

\end{abstract}





\maketitle

\section{Introduction}
On any given night in Canada, it is estimated that at least 35,000 individuals experience homelessness, and in a given year, at least 235,000 \cite{gaetz_homelessness_2016, homelesshub_about_2022}. These individuals may occupy unsheltered locations such as parks or sidewalks, or temporarily stay at emergency housing shelters. In addition to drop-in accommodation, emergency shelters provide long-term services to help individuals transition out of homelessness and into permanent housing programs. The front-line staff at these shelters, such as case workers, security staff, and day-to-day shelter coordinators, play a critical role in clients' lives as they build long-term relationships with clients, support them in achieving their goals, help maintain order within the shelter, and otherwise ensure the shelter functions effectively \cite{homelesshub_about_2022}. 

Client data plays an important role when staff devise strategies for how to best serve individuals experiencing homelessness, as it is common for front-line staff to both record and engage with a variety of data about each client \cite{thomas_principles_2020}. Data on one client can stretch back many years and include detailed historical information on the client such as building check-ins, notable interactions with staff or other clients, incident reports and progress toward becoming housed, among other information. 

Prior work has investigated how to harness this shelter data to develop predictive algorithms for assessing client vulnerability and assisting with evidence-informed program delivery \cite{kithulgoda_predictive_2022, rice_linking_2023, karusala_street-level_2019, chelmis_challenges_2021}. However, it can be problematic to integrate automated data tools, which may oversimplify client context, into staff decision-making processes \cite{karusala_street-level_2019, chelmis_challenges_2021, bopp_disempowered_2017, cronley_invisible_2022}. Therefore, we investigated design opportunities in decision-making contexts where staff make difficult decisions through group discussion and collaborative engagement with client data. Given the abundance of data in shelter databases, staff require a robust interface to holistically understand client context and make informed decisions about each client. At present, however, emergency shelters often lack the appropriate interfaces to support these aims and navigate this vast amount of data. 

We address this challenge by undertaking an ongoing co-design collaboration with staff at the Calgary Drop-In Centre -- the largest emergency housing shelter in Calgary, Canada, with over 500 beds in their downtown location. Working closely with 11 front-line staff members over a period of five months in 2022, we conducted a series of one-on-one semi-structured interviews, observational studies, and co-design sessions to investigate how staff at the Drop-In Centre interact with recorded client data during group discussions to make decisions about \textit{client barring} -- temporarily disallowing a client from accessing shelter services following a harmful incident. In this paper, we describe our continuing collaboration with front-line staff in the design, implementation, and ongoing deployment of a data-navigation interface. We highlight key characteristics in the relationship between staff and client data, and suggest how designers may integrate these factors into the design of data-navigation interfaces at shelters. 

\section{Background}
We situate our work along two areas of prior literature. First, we identify a gap in the technology that has thus far been designed to support homeless populations. Most literature on technology in this area has focused on assisting clients -- the individuals experiencing homelessness -- directly. Such work has explored a variety of areas, including the use of mobile devices and personal digital artifacts in homeless populations \cite{woelfer_homeless_2011, woelfer_improving_2011, mohan_food-availability_2019}, perceptions of technology among homeless individuals \cite{le_dantec_designs_2008}, and how mapping an individual's journey through homelessness can inform how technology may best assist them \cite{chandra_critical_2021}. In contrast, our work assists clients \textit{indirectly}, by supporting the front-line staff who help transition clients out of homelessness. Limited work has adopted this lens to investigate how shelter technologies can support those experiencing homelessness. One notable exception is the work of Le Dantec et al. \cite{le_dantec_tale_2010, le_dantec_publics_2011}, who co-designed a message board system in a shelter for homeless mothers to enhance client-staff communication. We similarly used participatory methods to co-design a tool for front-line use. However, the current work differs from Le Dantec et al.'s as we focus on the design of an interface to support staff in engaging with client data when making high-stakes decisions about clients.

The second area of work that we build on is the use of data tools in housing shelters. Shelters routinely collect and store an abundance of data on each client and their interactions with the shelter. Much prior work has investigated how to analyse this client data to understand homelessness at the population-level. For example, shelter data can be analysed to identify broad trends in homelessness \cite{brush_data_2016, thomas_principles_2020, culhane_potential_2016}, and to develop machine-learning and predictive modelling algorithms to identify factors correlated with or contributing to homelessness \cite{messier_predicting_2022, tabar_identifying_2020, richard_validation_2019, hong_applications_2018}. 

In addition to research at the population-level, some work has investigated how to integrate data tools into program delivery within individual shelters (e.g. \cite{kithulgoda_predictive_2022, rice_linking_2023, karusala_street-level_2019, chelmis_challenges_2021}). Such work looks at data tools that use predictive modelling algorithms to assess the vulnerability of each client. Vulnerability scores from these tools can inform decisions about how to allocate limited resources to clients who are most in-need \cite{rice_linking_2023}. Automated data tools can provide an evidence-informed basis for making decisions about how to best serve clients. There is a demand for these tools in shelters as they allow for staff to make decisions efficiently, and feel as though they are not clouded by personal biases \cite{karusala_street-level_2019, rice_linking_2023}. However, although housing shelters strive for the ideal of data-driven practices, in reality, the use of automated data tools to streamline human decision-making can oversimplify client context and conflict with the human-centered mission of these organizations \cite{karusala_street-level_2019, chelmis_challenges_2021, bopp_disempowered_2017, cronley_invisible_2022}. When using these tools, staff often distrust the tool's assessments and will look through client data manually, to form their own opinion of client vulnerability \cite{karusala_street-level_2019}. Indeed, the most widely-used vulnerability-assessment tool has recently been phased out due to controversies around its ability to capture important societal context surrounding client vulnerability \cite{cronley_invisible_2022, shinn_moving_2022, orgcode_message_2022}. 


Evidently, there are many limitations to using automated data tools for decision-making at shelters. 
Given that staff will often sift through client data themselves, we focus our work on designing for situations where staff engage directly with client data when making high-stakes decisions via group discussions. Through an ongoing collaboration with a local housing shelter, we designed and deployed a novel data-navigation interface that staff are currently using with live client data during weekly boardroom meetings. In this paper, we share preliminary findings regarding how shelter staff interact with recorded client data, and how data-navigation tools can be designed to support this interaction.

\section{Co-Design Context}
Staff at the Drop-In Centre manage an abundance of data when seeking to understand each client's circumstances. Staff need a usable and robust interface for navigating through this large volume of data. However, the primary staff-facing user interface at the Drop-In Centre does not provide a comprehensive navigational flow to explore client data, and is not tailored to the unique goals that a staff member may have when engaging with client data. Consequently, understanding client context through recorded client data is a time-consuming process, and especially detrimental if it impacts time-sensitive decisions to help mobilize clients' transitions out of homelessness or ensure the safety of the shelter.

\subsection{Client Barring}
Due to the nature of this work, Drop-In Centre staff must occasionally place a ``bar'' on clients, disallowing them from entering the shelter for a specified period of time -- a troubling experience for all. Bars are placed if a client has broken building guidelines, for example by endangering the safety of staff or other clients. The \textit{category} of a bar increases with the severity of the incident. Category 1 bars are equivalent to asking the client to go for a walk before re-entering the building. Category 4 bars may last over a month and are placed after very high-severity incidents. Staff sometimes grant clients ``compassionate override'' due to extenuating circumstances such as exceptionally cold weather. While bars are critical to fostering the overall safety of the shelter, they are highly distressing for barred clients as they inhibit access to shelter services. Therefore, the Drop-In Centre has a Bar Review Committee (BRC) who reviews bars in weekly meetings. During these BRC meetings, interpreting and discussing client data plays a key role. 

Members of the BRC have the difficult task of deciding how to move forward with clients who have received high-category bars. Decisions may include changing the category of a bar, replacing a bar with a \textit{condition of entry} such as having a conversation with a supervisor before reentering the building, or lifting a bar altogether. These decisions require balancing the barred client's need for shelter, with the shelter staff and other clients' needs for safety. During weekly meetings, BRC members review a wide variety of recorded client data in an attempt to gain a comprehensive overview of the incident and of the client's context. The goal is to make a decision that is fair to all.

\subsection{Client Data}
Over the course of a year, the Drop-In Centre serves over 4,000 unique individuals, and some clients may continue to interact with the shelter for more than a decade. Staff at the Drop-In Centre record qualitative and quantitative data on the interactions that clients have with the shelter. Data is organized into a database containing over 50 interrelated tables. As of March 2022, the Drop-In Centre database contained information on over 45,000 clients. Described below is a subset of relevant tables used in the implementation of the novel data-navigation tool.

The \textit{Clients} table contains general demographic information for each unique client. A new row is created for every client when they first interact with the shelter. The \textit{Building Check-Ins} table contains data regarding every instance of a client checking into the building. Each row contains, for example, the time of the check-in, the checked-in client, and the staff member who checked-in the client. The \textit{Bars} table contains data on each bar placed on clients at the shelter. It includes data on the client who received the bar, the category and duration of the bar, and a brief one-line description of why the bar was placed (such as ``Unprovoked harm to another client''). The \textit{Logs} table is the most commonly-used table. For nearly every encounter a staff member has with a client, they write a new log entry describing the event. Logs may be written, for example, when a staff member witnesses an altercation between clients, when a client's housing status changes, when a client expresses enthusiasm for a particular service, or after any other notable event that staff believe should be recorded. 

Prior to the deployment of their new interface, BRC members interacted with the database via Guestbook, a graphical user interface implemented using Microsoft Dynamics. Guestbook displays numerous tables containing all recorded client data. The limitation of this method of interacting with client data is that Guestbook is used for all possible data interactions, regardless of the specific purpose for interacting with the data. This lack of customization leads to a highly cluttered and time-consuming interface.


\section{Methods}
We collaborated with a cohort of 11 staff members at the Drop-In Centre. The cohort included representation from multiple teams: senior managers, shelter managers, shelter coordinators, security staff, impact analysts, and staff members responsible for placing clients into housing programs. We conducted semi-structured interviews, observed Bar Review Committee meetings, and led co-design sessions, with the aim of characterising how staff engage with client data when making decisions about bars, and developing a tool to support this process.

\subsection{Semi-Structured Interviews} 
We conducted semi-structured interviews with five staff members involved in the barring process. Interviewees included two shelter managers, one senior manager, one senior shelter coordinator, and one coordinator of security. Interviews were 30-60 minutes long and led by one researcher with one other researcher present. Since barring can be a sensitive topic among staff, we opted not to video or audio-record interviews. This decision helped to minimize the sense of formality during interviews and reduced the social gap between researcher and interviewee. Both researchers thus took extensive notes throughout each interview. Interviews were composed of two sections. First, "day-in-the-life" questions investigated the participant's role at the Drop-In Centre, their relationship with technology, and what they enjoy and find challenging about their work. The purpose of these questions was to understand how a new technology may fit their existing workflow. The second set of questions were specific to bars and investigated which data participants are most interested in when reviewing bars, how they typically interact with data using Guestbook, and how they would ideally like this interaction to look.

\subsection{Bar Review Committee Meeting Observations}
The Bar Review Committee includes staff from various departments at the Drop-In Centre who meet once a week for two hours. During BRC meetings, members discuss barred clients and decide how to move forward with them. Two researchers observed a BRC meeting prior to designing the new interface. The meeting included three members of our cohort -- one member of the housing team, one shelter manager, and one shelter coordinator -- as well as four additional staff members who were not actively involved in the co-design process. Researchers took field notes throughout the meeting to record how participants interacted with Guestbook while discussing each bar. Notes included detailed accounts of the navigational flow through the data tables, what questions were asked about clients and how questions were answered, and where points of friction were when using Guestbook. After the first iteration of the new tool was deployed, a researcher observed another BRC meeting where members used the new tool to facilitate their discussion. Similar notes were taken during this second meeting. These observations provided direct insight into how staff engage with data and illuminated how the design could be modified to better support BRC discussions. 


\subsection{Co-Design Session and Iterative Deployment}
\begin{figure}[ht]
  \centering \includegraphics[width=\linewidth]{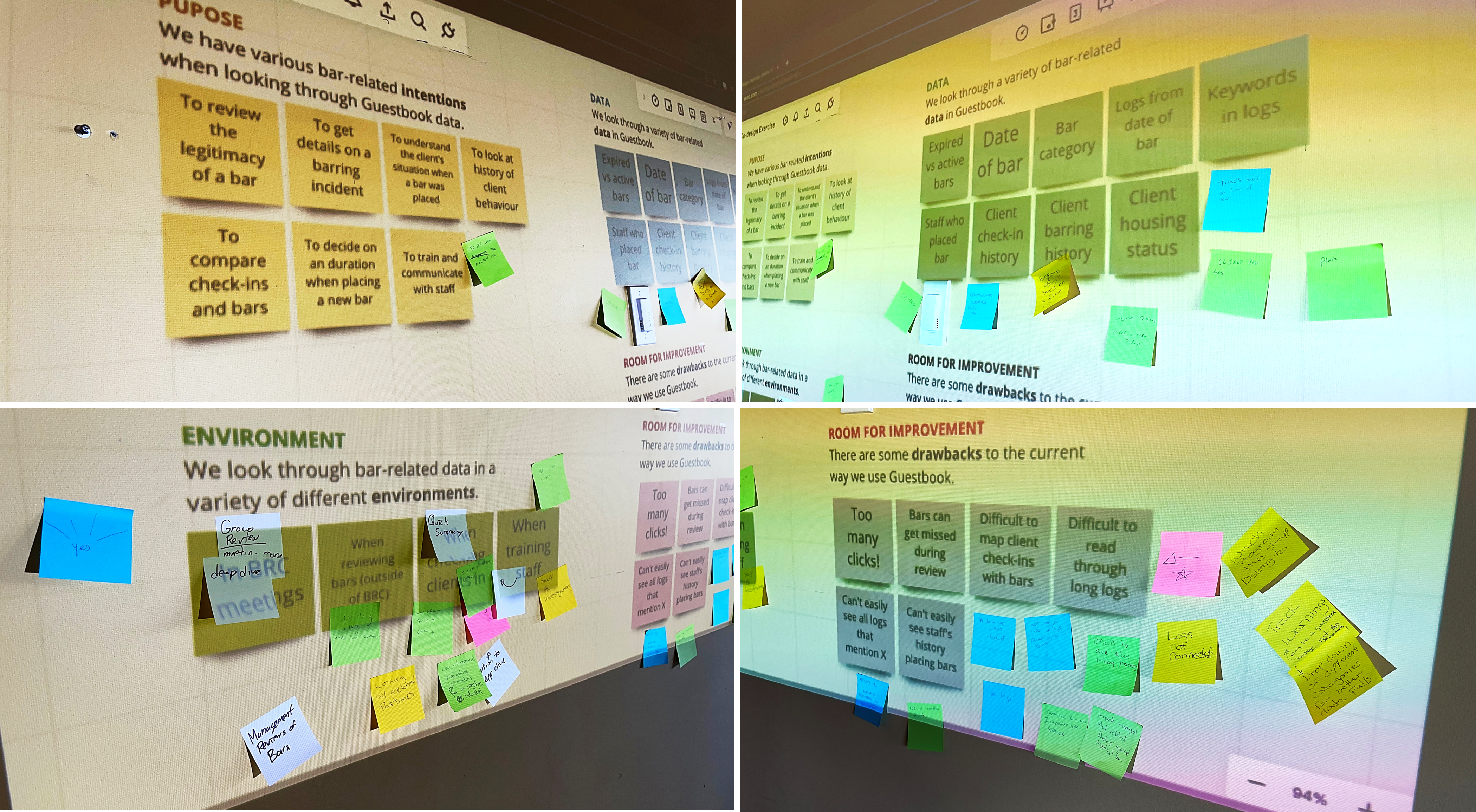}
  \caption{Four themes (purpose, environment, data, and room for improvement) with corresponding research notes were projected onto a wall. Participants added 30 sticky notes to the wall based on their own interpretations of each theme.}
  \label{fig:WallWalk_Photos}
  \Description{Four photos depicting the themes "Purpose," "Data," "Environment," and "Room for Improvement" projected onto a wall, with multiple multi-coloured sticky notes stuck to the wall clustered within each theme.}
\end{figure}

Thematic analysis of notes from interviews and the first BRC observation revealed four preliminary emergent themes which characterised staff members' interaction with client data: the data, the purpose, the environment, and room for improvement. These themes informed the co-design session that occurred after the first BRC observation. The co-design session included the full cohort of staff, as well as three members of the IT team who had additional insight into how Guestbook is used. The session began with an affinity diagramming exercise and led into a sketching exercise. For the affinity diagramming exercise, the four emergent themes were projected onto a wall with corresponding research notes from interviews and the BRC observation. Staff then participated in an adaptation of Holtzblatt's Immersive Wall Walk exercise which encourages participants to think as designers by dynamically and creatively interacting with design themes \cite{holtzblatt_10_2017}. Participants interpreted the themes and added new (physical) notes onto the (projected) diagram, based on their own reflections on how they engage with client data (shown in Figure \ref{fig:WallWalk_Photos}). The communicated purpose of this exercise was for staff and researchers to converge on a shared understanding of how staff interact with data in the context of bars.

\begin{figure}[ht]
  \centering \includegraphics[width=\linewidth]{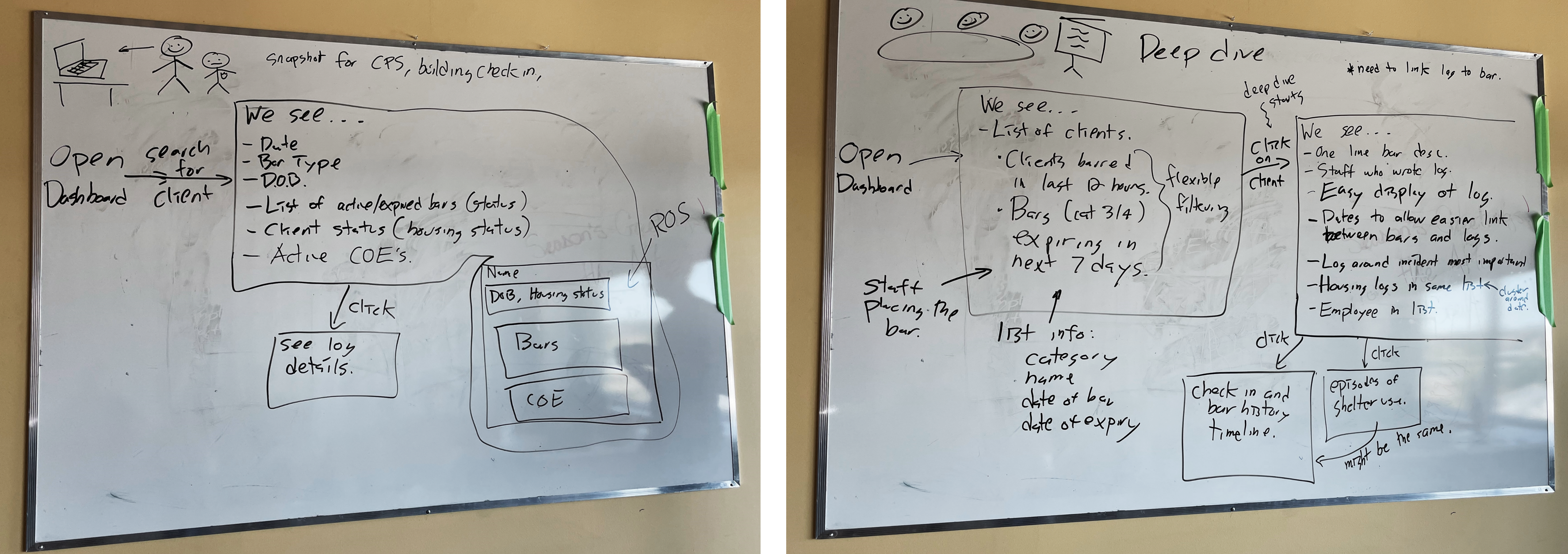}
  \caption{Researchers and participants collaboratively developed sketches for the "Quick Summary" (left) and "Deep Dive" (right) data-navigation contexts.}
  \label{fig:Sketches_Photos}
  \Description{Two photos of a white board, containing sketches of a "Quick Summary" concept and a "Deep Dive" concept. Both sketches indicate of what types data the user may see in each potential interface design and how they will interact with with interface to see more data.}
\end{figure}

After the affinity diagramming exercise, two researchers led a group discussion where notes from the \textit{environment} theme were clustered based on which data is accessed in each environment, and the typical purpose for accessing such data in each environment, leading to two overarching types of navigation: ``Quick Summary'' and ``Deep-Dive.'' Participants and researchers then collaboratively brainstormed low-fidelity sketches to capture how the new data-navigation tool may look for each type of navigation. Figure \ref{fig:Sketches_Photos} depicts sketches of the Quick Summary and Deep Dive navigation concepts. Each sketch contained descriptions of what data should be shown, and how staff may interact with the interface to view different data. For the purpose of facilitating decision-making during BRC discussions, the ``Deep-Dive'' concept was preferred. After designing an initial prototype of the data-navigation interface, the cohort was invited to informal feedback and brainstorming sessions to observe how the concepts from the co-design exercise translated to a high-fidelity prototype. Notes were taken to capture initial thoughts on the design, suggested modifications, and additional reflections on how the interface will be used. This process was then repeated for the second iteration of the interface, which is currently in-use with live client data during weekly BRC meetings at the Drop-In Centre. The original and current versions of the design are shown in the Appendix.

Using these co-design methods, we aimed to explore the relationship between front-line staff and recorded client data. Analysing this relationship via such a hands-on project led to rich, real-world findings. However, findings from this study may be limited in their specificity to the North American culture surrounding homelessness and may not be generalizable to other regions and cultures.

\section{Preliminary Findings: Recognizing Bias in Client Data and Striving for Compassionate Decision-Making}

Analysis from semi-structured interviews, Bar Review Committee meeting observations, and co-design sessions led us to identify several emergent themes in how shelter staff engage with data to understand client context. Below, we detail the most prominent characteristic of staff members' interaction with client data: staff are highly familiar with the ``messiness'' of recorded client data, and this familiarity gives rise to careful discussions with the intent of compassionate decision-making.

Client data can be ``messy'' for a variety of reasons. All interviewees explained that reviewing bars is difficult due to the subjective nature of recorded client data, and the need to consider all potential biases that could have influenced data-entry. For example, a coordinator described the subjectivity of placing bars: \textit{``there can be bias depending on how sensitive the staff person is to the event... some staff are very hurt by yelling and want long bars for things that aren't really justified.''} Similarly, a manager explained that \textit{``a Category 4 bar may be due to a triggered staff... some staff might be barring a lot more than others so we have to look at that.''} These issues also arise when staff members write logs. Logs provide a direct account of an incident, which is important to read when reviewing bars. However, whether or not certain information is included in log notes is dependant on a) what the writer happened to remember or notice about a client encounter, and b) what the writer considered worth recording about the event. 
It was clear from our analysis that such messiness also extends beyond qualitative data, such as log notes, and influences certain \textit{quantitative} metrics as well. For example, client ``acuity'' scores are meant to indicate the client's level of vulnerability, and are calculated from a variety of client data including the number of bars a client has received. One staff member who is responsible for housing triage explained that they do not consider these scores to be particularly informative due to the bias inherent in all recorded data. In fact, they sometimes disregard these scores altogether and instead read through clients' log histories to obtain their own understanding of the client. 

Staff members' collective familiarity with the bias in client data informed how they treated such data during BRC discussions. These discussions involved iteratively inquiring about each client being discussed. For example, a BRC member may ask what the reason was for the client's previous bar or about their history of shelter-use. These questions would then be answered by navigating through the data in search of an explanation, and then interpreting and discussing the data. Notably, the information found in the data rarely led to a definitive answer. Instead, the many subjective nuances of the data invited further discussion and encouraged staff to look beyond the data. Discussions would rely on personal interpretations of the data, social work expertise or knowledge of societal context, and sharing of personal client-facing experiences. 

Analysis of co-design sessions further revealed that client data is treated as a conversation-starter rather than a definitive depiction of clients' circumstances. This theme was exemplified by a conversation that occurred during a co-design session about the need for the interface to encourage compassionate decision-making. A senior manager disagreed with an impact analyst's idea to include a bar-count in the interface for each client. The senior manager articulated that such a number would stifle BRC discussions by making it \textit{``too easy to judge clients based only on bars''} and would dissuade BRC members from \textit{``digging deep''} so that they can adequately understand each client. ``Digging deep'' in this context referred to personally interpreting and discussing qualitative data to come to a more carefully-considered decision about each client.

Findings from interview analyses additionally explain what it means to look beyond the data to gain a holistic understanding of each client. 
During their interview, a coordinator explained that BRC members occasionally discuss whether or not client behaviour was ``intentional.'' Such discussions are based on subjective interpretations of logs as well as knowledge of the societal context in which clients live. For example, a BRC member with social work expertise may provide more nuanced insight into the degree of ``intention'' behind a harmful behaviour compared to the log entry that gives a direct account of the incident. A shelter manager additionally explained that members of the BRC often ``advocate'' for particular clients, based on client-facing experience and their personal understanding of the client's circumstances. Therefore, it seemed that unrecorded information about clients played a critical role in decision-making, and even held more value than what was depicted in the data.  

\section{Discussion}
Preliminary findings from analysis of our co-design study have revealed a complex interaction between client data and staff decision-making processes. The key challenge when designing in this space is to consider how to handle such ``messy'' data to best support evidence-based decision-making in high-stakes contexts. One solution may be to reduce ``messiness'' and provide users with more concrete information, perhaps by including keyword counts from aggregates of logs, or foregrounding data that is less likely to have strong biases. For example, an interface could de-emphasize logs that were written by a staff member who seems to disproportionately place high-category bars. However, preliminary findings demonstrate that attempting to minimize the subjective nuances in recorded client data is not conducive to ongoing discussions and further inquiry into client context. Instead, this approach may preemptively end conversations by presenting seemingly definitive data about clients. Alternatively, staff could aim to standardize data collection to minimize bias. Managers and supervisors at the Drop-In Centre have, in fact, worked towards optimizing the integrity of recorded client data by encouraging staff to standardize language when writing logs and reasons for bars. However, when aiming to make compassionate decisions through discussion, the BRC still does not trust the data to provide a true representation of the client.

Notably, recent work has scrutinized the notion that data can be neutral or objective. Such works instead emphasize that data is always collected \textit{by someone} and \textit{for some purpose}, and is, therefore, always an intentionally simplified and reductive model of reality \cite{dignazio_creative_2017, sanches_diffraction--action_2022}. In the context of pedagogy, D'Ignazio \cite{dignazio_creative_2017} urges educators to cultivate a skepticism of ``raw'' data to empower their learners to question the seemingly objective nature of data. Similarly, in the context of personal informatics, Sanches et al. \cite{sanches_diffraction--action_2022} proposed that, since data is ``entangled'' with societal context, designers may benefit from ``holding space'' for messy and ambiguous data that requires active and effortful interpretation on the part of the user. Our findings build upon this prior work, revealing similar effects in the context of a housing shelter where staff attempt to understand clients through recorded data.

Furthermore, prior work highlights that when shelter staff use automated data tools for decision-making, they often feel as though the ``human-touch'' is missing from their work \cite{karusala_street-level_2019}, leading them to instead look directly at client data to make a their own assessments of clients' needs. In the present work, we investigated a situation where staff are, in fact, engaging directly with recorded client data to make decisions. Still, there was a sense of distrust in the data as it only provided a starting point for decision-making, and staff members consistently looked to greater context before feeling comfortable coming to a decision. When designing a tool to support compassionate, discussion-based decision-making in a high-stakes context, then, it may be beneficial for designers to avoid presenting the data as though it is neutral and objective. Instead, supporting decision-making in this context may necessitate designs that hint at context outside the data itself and encourage staff to discuss what is missing, rather than encouraging a seemingly-definitive understanding of client circumstances.

\section{Conclusion and Future Work}
We presented the initial results of an ongoing co-design and deployment of a data-navigation interface at the largest emergency housing shelter in Calgary, Canada. Our goal for the interface is to support shelter staff when faced with the challenging task of deciding how to move forward with barred clients. Such decisions involve reviewing and discussing recorded client data to understand each client's unique circumstances. Our work contributes to the body of work on technology and data tools to support housing shelters and individuals experiencing homelessness. We highlighted a series of insights: staff recognize the inherent bias in client data; client data can only depict part of a client's story; and these facets of the staff-data relationship inform how staff prioritize a compassion-oriented understanding of the client. We will continue to collaborate with staff at the Drop-In Centre to design and deploy data tools that align with their values. Future directions of this research include understanding and designing for additional decision-making contexts within the shelter, such as decisions regarding housing placements for clients and decisions about staff training procedures. We emphasize the importance of close collaborations between designers and shelter staff so that data tools may better fit into and support the relationship between staff and client data. 


\begin{acks}
The authors would like to gratefully acknowledge the support of Making the Shift, the Calgary Drop-In Centre, and the Government of Alberta. This study is based in part on data provided by Alberta Seniors, Community and Social Services. The interpretation and conclusions contained herein are those of the researchers and do not necessarily represent the views of the Government of Alberta. Neither the Government of Alberta nor Alberta Seniors, Community and Social Services express any opinion related to this study.
\end{acks}


\bibliographystyle{ACM-Reference-Format}
\bibliography{DI_LBW_Arxiv.bib}

\appendix
\section{APPENDIX: Interface Design}

\begin{figure}[H]
  \centering \includegraphics[width=\linewidth]{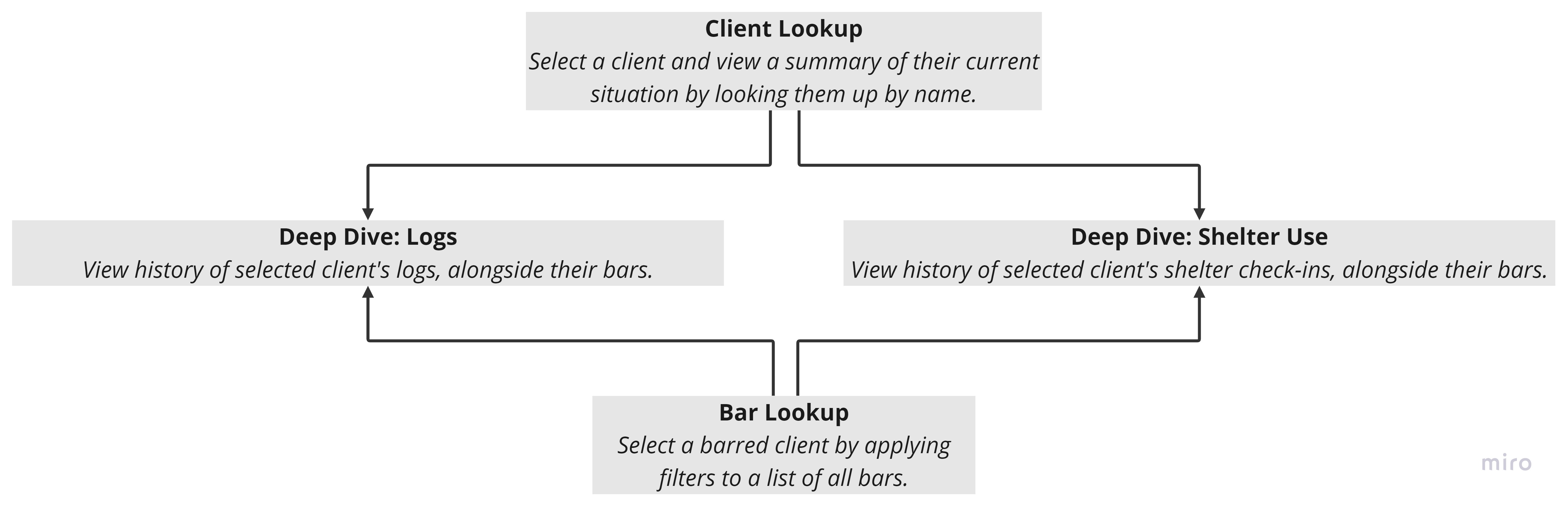}
  \caption{[First Version] The navigational flow from the Client Lookup and Bar Lookup pages to the Logs and Shelter Use pages.}
  \Description{A flow chart with four boxes: 1) Client Lookup, 2) Bar Lookup, 3) Deep Dive: Logs, and 4) Deep Dive: Shelter Use. Two arrows point from Box 1 to Box 3 Box 4. Another two arrows point from Box 2 to Box 3 and Box 4.}
  \label{fig:NavigationDiagram}
\end{figure}

\begin{figure}[H]
  \centering \includegraphics[width=\linewidth]{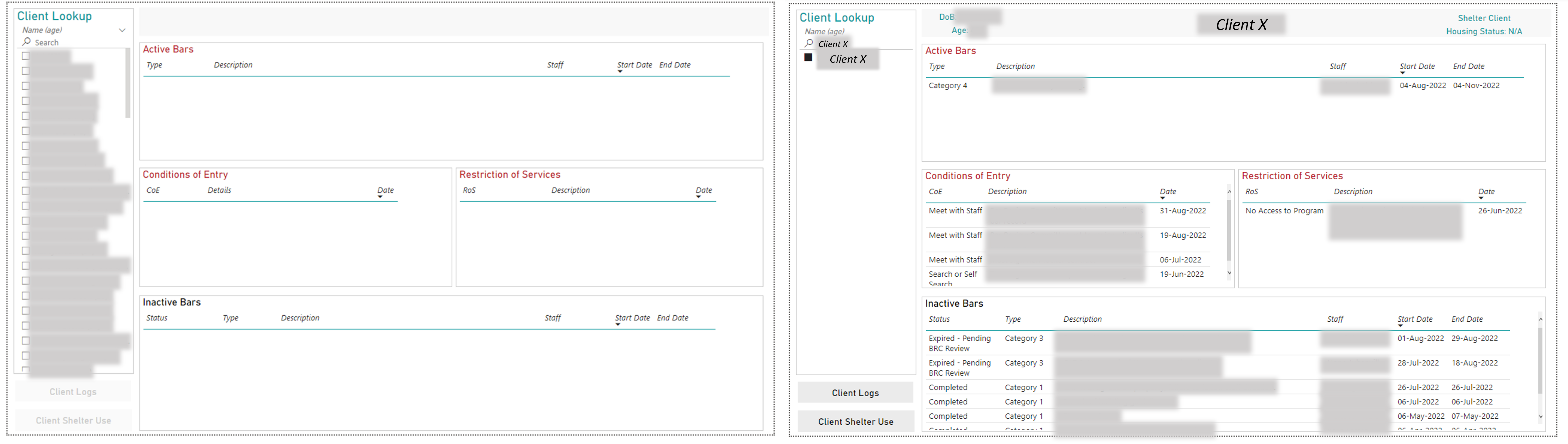}
  \caption{[First Version] The Client Lookup page with no client searched for (left), and with a client searched for and selected (right). All identifying information about Drop-In Centre clients and staff are anonymized using grey boxes.}
  \Description{The Client Lookup page contains a panel to search for clients by name, and four additional panels: Active Bars, Restriction of Services, Conditions of Entry, and Inactive Bars. When no client is selected, these four panels are empty. When a client is selected, these four panels populate with data regarding that client and two buttons become enabled: Client Logs and Client Shelter Use.}
  \label{fig:ClientLookupScreenshots}
\end{figure}

\begin{figure}[H]
  \centering \includegraphics[width=\linewidth]{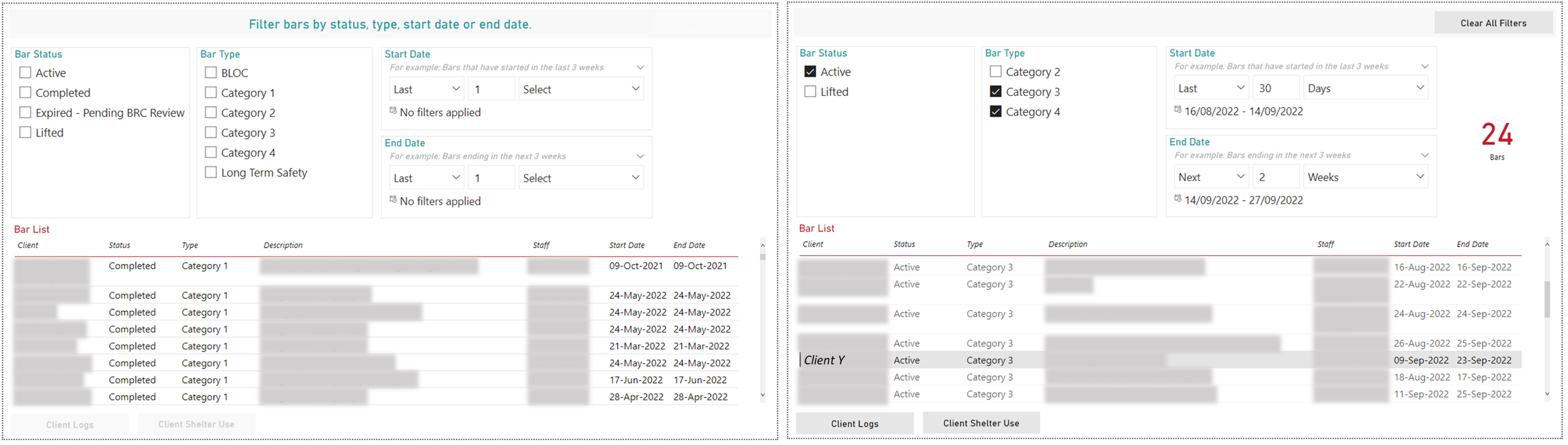}
  \caption{[First Version] The Bar Lookup page with no filters applied (left), and with filters applied and a barred client selected (right). All identifying information about Drop-In Centre clients and staff are anonymized using grey boxes.}
  \Description{The Bar Lookup Page contains a formatted table with every bar that has been placed. There are filters for Bar Status (Active, Completed, Expired, or Lifted), Bar Type (BLOC, Category 1, Category 2, Category 3, Category 4, or Long Term Safety), Start Date of the bar, and End Date of the bar. When filters are applied, a number of how many resulting bars there are in displayed. When a specific bar is selected, two buttons become enabled: Client Logs and Client Shelter Use.}
  \label{fig:BarLookupScreenshots}
\end{figure}

\begin{figure}[H]
  \centering \includegraphics[width=\linewidth]{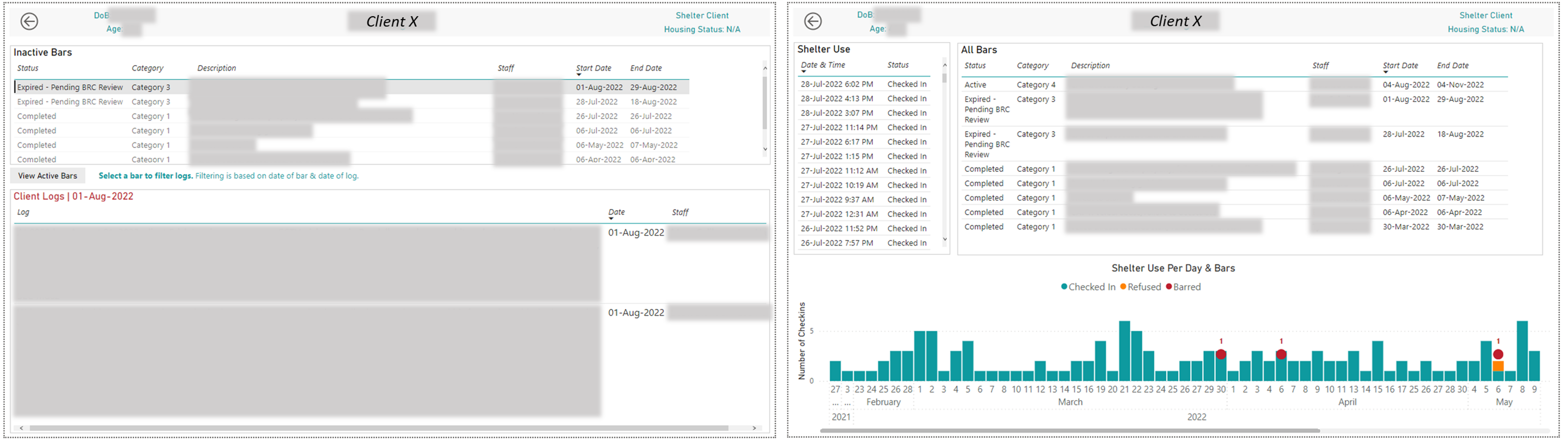}
  \caption{[First Version] The Client Logs page (left) and the Client Shelter Use page (right). All identifying information about Drop-In Centre clients and staff are anonymized using grey boxes.}
  \Description{The Client Logs page contains a table with all written logs about the selected client, as well as a table with their bars. The user can select a bar to filter the logs table by corresponding date. The Shelter Use page contains a stacked column chart with dates on the x-axis and Number of Encounters on the y-axis. Green column segments indicate building check-ins, and orange indicates refusals. Red circles are incorporated into the bar chart to indicate when the client has been barred in relation to their building check-ins.}
  \label{fig:DeepDiveScreenshots}
\end{figure}

An overview of the navigational flow of the original interface design is shown in Figure \ref{fig:NavigationDiagram}. The navigation of the interface began with either the Client Lookup or Bar Lookup page. The Client Lookup page (Figure \ref{fig:ClientLookupScreenshots}) allowed staff to search for a particular client by name. As indicated in the the Quick Summary sketch, once the staff member selected a client, a summary of the client's bars, conditions of entry, and restrictions of services was displayed, along with their date of birth, client type, and housing status. The Bar Lookup page (Figure \ref{fig:BarLookupScreenshots}), based on the first frame in the Deep Dive sketch, contained a history of all bars at the shelter. The staff member could filter this list to select a particular barred client. A ``Deep Dive'' could then be initiated from either the Client or Bar Lookup page. The Deep Dive pages (Figure \ref{fig:DeepDiveScreenshots}) included a log-history page and a shelter-use page.

\begin{figure}[H]
  \centering \includegraphics[width=\linewidth]{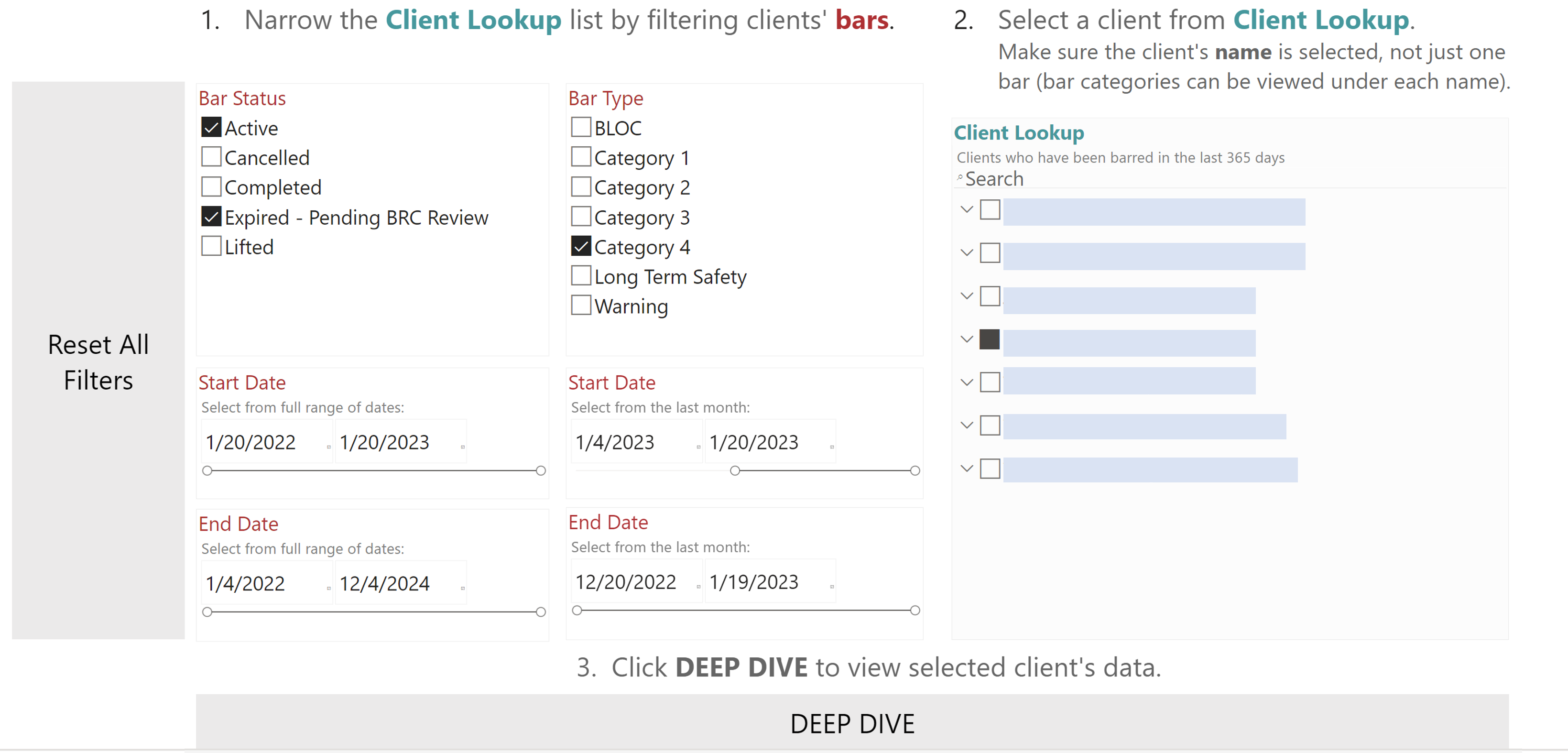}
  \caption{[Current Version] The Client Lookup page, with filters and a client selected. All identifying information about Drop-In Centre clients and staff are anonymized using blue boxes.}
  \Description{The current Client Lookup Page contains a panel to search for a client, as well as filters to filter the client list based on their bars. Once a client is selected, a button becomes enabled: Deep Dive.}
  \label{fig:Lookup2}
\end{figure}

\begin{figure}[H]
  \centering \includegraphics[width=\linewidth]{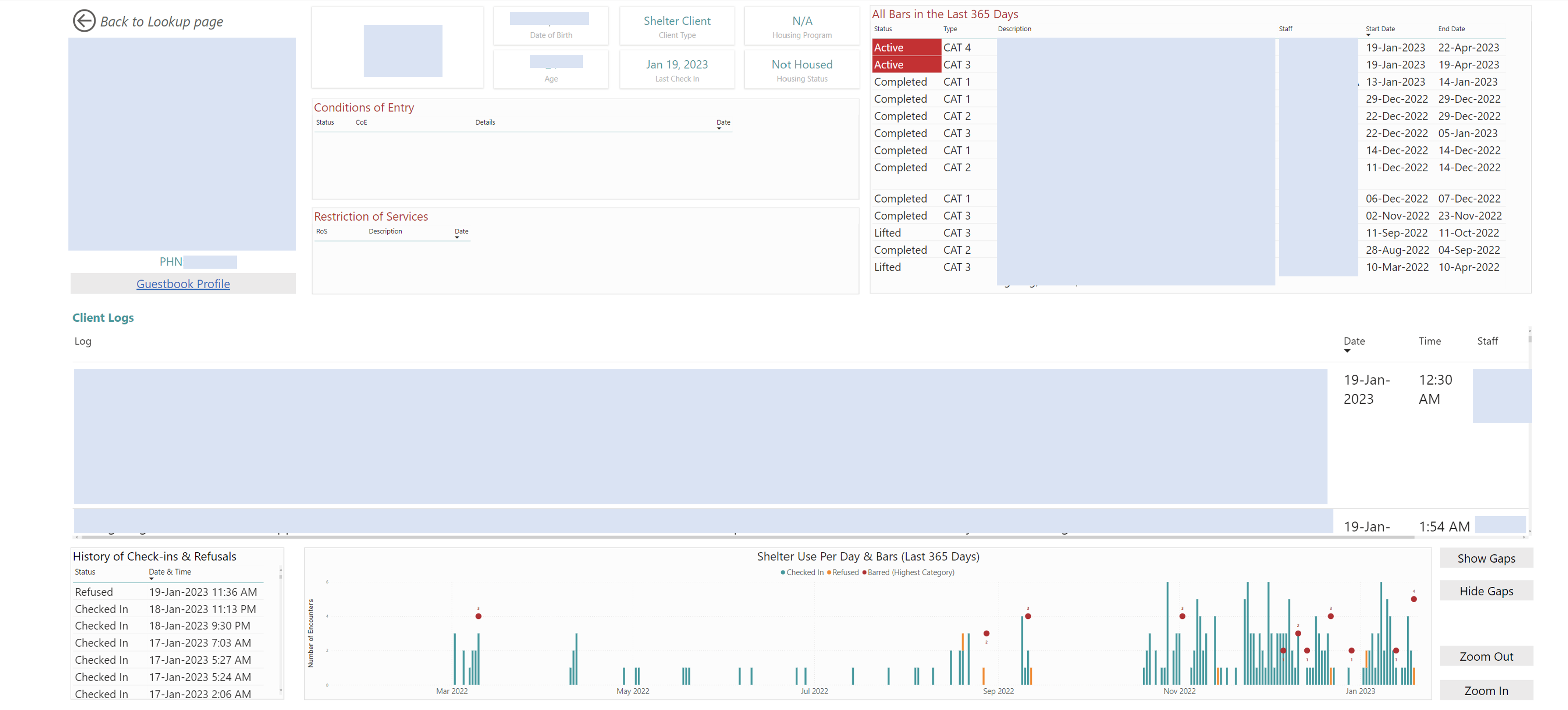}
  \caption{[Current Version] The Deep Dive page, navigated to from the Deep Dive button on the Client Lookup Page. Users can select a bar, CoE, RoS, or date on the shelter-use chart to filter client logs based on date. All identifying information about Drop-In Centre clients and staff are anonymized using blue boxes.}
  \Description{The current Deep Dive page contains all relevant data for the selected client: client photo, health number, demographic information, bars (with active bars highlighted in red), logs, and shelter use. In the top left corner there is a button to go back to the Lookup Page.}
  \label{fig:Deepdive2}
\end{figure}

Staff feedback on the first prototype emphasized the need to have all pertinent data ``in one place.'' Thus, the second design was condensed to only include a Look Up page (Figure \ref{fig:Lookup2}) and a Deep Dive page (Figure \ref{fig:Deepdive2}). Throughout the BRC meeting, members alternate between looking up a barred client either by name or by adjusting the filters, and then interacting with the Deep Dive page. In the Deep Dive page, staff members can select the date of any bar, CoE, RoS, or building check-in, to filter the logs and only see entries from the selected date. Staff can also adjust the view of the shelter-use chart.

\end{document}